\documentclass[doublecol]{epl2} 

\usepackage[tbtags]{amsmath}

\title{Corrections to the thermodynamic quantities of Bose system by the generalized uncertainty principle}
\shorttitle{Corrections to the thermodynamic quantities of Bose system by the GUP} 

\author{Jun-Xian Li\inst{1}\thanks{E-mail: \email{jxli\_2020@163.com}} \and Jing-Yi Zhang\inst{1}\thanks{E-mail: \email{zhangjy@gzhu.edu.cn(corresponding author)}}}
\shortauthor{J.X. Li and J.Y. Zhang}

\institute{                    
  \inst{1} Centre for Astrophysics, Guangzhou University - Guangzhou 510006, China
}

\abstract{
This paper investigated the Bose system in a spherical shell close to the black hole horizon. Several thermodynamic quantities of the Bose system are derived, which are different from those in the flat spacetime, by introducing the generalized uncertainty principle (GUP) into the grand partition function of statistical mechanics. The internal energy and the pressure of the Bose system appear to have a correction term of $T^6$, while the entropy has a $T^5$ correction term where both the coefficients are functions of the spacetime component $g_{00}$ and the brick-wall model parameter $\epsilon$. Taking the Schwarzschild black hole as an example, the physical quantities of the shell such as temperature, pressure and entropy are calculated for the final stage of black hole radiation.
}

\begin{document}

\maketitle

\section{Introduction}

In 1973, inspired by the black hole's area theorem\cite{hawking1971gravitational}, Bekenstein\cite{bekenstein2020black} noticed the striking similarities between black hole mechanics and thermodynamics and pointed out that a black hole is a thermodynamic system. Hawking\cite{hawking1975particle}, who was initially opposed to the idea, proved that black holes have thermal radiation by means of the method of quantum field theory in curved spacetime, showing that the four laws of black hole mechanics are essentially the four laws of conventional thermodynamics. Since the black hole radiation is purely thermal, it is natural that there will be a concomitant loss of information when a black hole radiates all its mass, known as the black hole information puzzle. At the same time, the existence of the singularity in black holes is so incredible and incomprehensible that people begin to search for a quantum gravity theory that could accommodate both general relativity and quantum theory. As strong candidates for quantum gravity, studies\cite{veneziano1986stringy,GROSS1988407,yoneya1989interpretation,AMATI198941,KONISHI1990276,MAGGIORE199365,kempf1995hilbert,SCARDIGLI199939,adler1999gravity,ahluwalia2000wave} such as string theory and loop quantum gravity suggested that the Heisenberg uncertainty principle (HUP) should be modified when the gravitational effect is non-negligible. Then the counterpart is the generalized uncertainty principle (or the gravitational uncertainty principle)
\begin{equation}
    \Delta x \geq \frac{\hbar}{\Delta p} + \frac{\alpha}{\hbar} \Delta p,
\end{equation}
where $\alpha \sim G$. All these studies have shown that the GUP is a general conclusion which is independent of the specific gravitational quantization scheme. However, the deficiency of HUP is that it does not take the gravitational interaction into account. As previous studies of black hole thermodynamics have been using the HUP, some of the earlier results may need to be modified if the GUP replaces the HUP. Through a heuristic approach where the HUP prevents the total collapse of hydrogen atoms, R.J.Adler \etal \cite{adler2001generalized} utilised the GUP to prove that the complete evaporation of a black hole can be prevented by dynamics rather than by symmetry or quantum numbers. Many works \cite{adler2001generalized,Cust_dio_2003,CHEN2005233,MAZIASHVILI2006232,CHEN20151} have shown that considering the GUP would allow a black hole to evaporate to the point where there would be a remnant of what is considered to be one of the candidates for cold dark matter left at the end. Furthermore, A.J.M.Medved and E.C.Vagenas \cite{medved2004conceptual} obtained an unfolded quantum modified Bekenstein–Hawking entropy through the GUP, proving that the prefactor of the logarithmic term of the entropy can be represented by a parameter derived from a more fundamental theory. Numerous calculations \cite{medved2004conceptual,MYUNG2007393,LI2007207,kim2007entropy,nozari2007existence,Nouicer_2007,Kim_2008,PARK2008698,HAN2008121,nozari2012natural,He_2014,feng2016quantum} have used the GUP method to study black hole thermodynamics to derive corrections to the Hawking radiation spectrum and Bekenstein-Hawking entropy.

In 2014, Ref.\cite{He_2014} introduced the GUP into a shell with Bose gas to obtain a modified Stefan-Boltzmann law, which showed a $T^6$ correction term. Applying the modified law to the black hole radiation yields a black hole remnant with the Planck scale and a maximum radiation temperature at that time. In addition, the lifespan of a black hole is also modified. We believe some corrections will appear in the Bose system by considering the GUP. Ref.\cite{He_2014} did not further research the corrections to other thermodynamic quantities of the Bose system, so our work builds on this.

The remaining part of the paper proceeds as follows: The second part briefly describes the method for obtaining the energy-momentum relation for Bose gas in the Schwarzschild spacetime. In the third part, we derive the corrections to the thermodynamic quantities by means of the grand partition function. The fourth part contains a brief discussion and conclusion. The unit $G=c=1$ is adopted throughout this paper.

\section{Energy-momentum relation of particles in curved spacetime}
In order to study the thermodynamic properties of a black hole, 't Hooft \cite{THOOFT1985727} proposed the famous "brick wall" model in 1985. In this paper, we adopt this model and move the distant wall to the vicinity of the black hole horizon, so that a thin shell adjacent to the horizon is formed between the two walls. The thermal equilibrium quantum gas(Hawking radiation) within this shell near the Schwarzschild black hole horizon is selected as the research object. The Hawking radiation in the shell is
\begin{equation}
    N_\omega^2=\frac{1}{e^{\omega/k_B T} -1},
\end{equation}
and this spectrum is the particle number distribution law of the Bose system\cite{sannan1988heuristic}.
Since the shell is infinitely close to the horizon, a coordinate system which does not diverge at the horizon should be adopted. Introduce the following coordinate transformation into the Schwarzschild metric,
\begin{equation}
    t=t_s +2\sqrt{2Mr}+
    2M\ln\frac{\sqrt{r}-\sqrt{2M}}{\sqrt{r}+\sqrt{2M}},
\end{equation}
where $t_s$ is the Schwarzschild time. Then the Schwarzschild spacetime line element can be written as
\begin{equation}
    \upd s^2 =-f\upd t^2+2\sqrt{1-f}\upd t \upd r + \upd r^2 + r^2 \upd \theta^2 + r^2 sin^2\theta \upd \varphi^2,
\end{equation}
where $f=1-\frac{2M}{r}$, M for the black hole mass, $r_H=2M$ for the event horizon. It is a stationary coordinate system without divergence behaviour at the event horizon $r_H$, first proposed by Painlevé \cite{painleve1921mecanique}. We take a thin shell layer for discussion which is based on the brick-wall model at the horizon, whose inner surface is infinitely close to the horizon. The coordinate distance between the inner surface and the horizon is $\epsilon$ and the thickness of the shell is $\delta$. 

We consider that the shell is filled with Bose gas.
The Klein-Gordon equation satisfied by Bose particles is
\begin{equation}
\label{eq.KGeq}
    \frac{1}{\sqrt{-g}}\partial_\mu (\sqrt{-g}g^{\mu \nu} \partial_\nu \phi)=0.
\end{equation}
Use the Wentzel-Kramers-Brillouin (WKB) approximation and set the wave function as
\begin{equation}
\label{eq.wavfunc}
    \phi=e^{-i\omega t} e^{\frac{i}{\hbar}S(r,\theta,\varphi)}.
\end{equation}
Combining Eq.(\ref{eq.KGeq}) and Eq.(\ref{eq.wavfunc}) yields
\begin{equation}
    \hbar^2 \omega^2 + 2\sqrt{1-f}\hbar \omega p_r -fp_r^2 -\frac{1}{r^2} p_{\theta}^2 -\frac{1}{r^2 sin^2\theta} p_{\varphi}^2 =0,
\end{equation}
where $\omega=p_t,p_r=\frac{\partial S}{\partial r},p_\theta =\frac{\partial S}{\partial \theta}, p_\varphi =\frac{\partial S}{\partial \varphi}$. For a spherically symmetric system, only the radial motion is taken into consideration and then the energy-momentum relation for the boson can be found as
\begin{equation}
    p=\frac{(1\pm \sqrt{1-f}) \hbar \omega}{f}.
\end{equation}
Since the modulus of momentum cannot be negative, only the positive root is taken. Considering the shell is close to the horizon, $f \to 0$, therefore we have
\begin{equation}
\label{eq.EMrlt}
    p=\frac{2\hbar \omega}{f}.
\end{equation}

Then we obtained the energy-momentum relation for Bose gas in the vicinity of the event horizon.

\section{Corrections to the thermodynamic quantities of Bose system in a shell}

To understand the behaviour and thermodynamic properties of the Bose system in a shell, we will discuss it in a 6-dimensional phase space. By the GUP there is
\begin{equation}
    \Delta q_1 \Delta q_2 \Delta q_3 \Delta p_1 \Delta p_2 \Delta p_3 = \hbar^3 (1+\frac{\alpha}{\hbar^2} \Delta p^2)^3,
\end{equation}
where $\Delta q_i$ and $\Delta p_i$ represent the position and momentum uncertainty, respectively. In a 6-dimensional phase space, the degeneracy $\omega_l$ is equal to the volume $\upd x \upd y \upd z \upd p_x \upd p_y \upd p_z$ divided by the uncertainty, i.e.
\begin{equation}
    \omega_l =\frac{\upd x \upd y \upd z \upd p_x \upd p_y \upd p_z}{\hbar^3 (1+\frac{\alpha}{\hbar^2} p^2)^3}.
\end{equation}
The above equation is the expression in a Cartesian coordinate system, it can be rewritten in the spherically symmetric coordinate as
\begin{equation}
    \omega_l =\frac{(4\pi )^2r^2 p^2 \upd r \upd p }{\hbar^3 (1+\frac{\alpha}{\hbar^2} p^2)^3}.
\end{equation}

For a Bose system in a shell, thermodynamic quantities can be derived from the partition function in statistical theory. Introduce a grand partition function as follow
\begin{eqnarray}
\begin{aligned}
    \ln \Xi & =-\sum_{l} \omega_{l} \ln(1-e^{-\beta \varepsilon_l})\\
    & \simeq -\int \frac{(4\pi )^2r^2 p^2 \ln (1-e^{-\beta \varepsilon_l}) }{\hbar^3 (1+\frac{\alpha}{\hbar^2} p^2)^3}\upd r \upd p.
\end{aligned}
\end{eqnarray}
Now, substituting the energy-momentum relation (\ref{eq.EMrlt}) into the above equation gives
\begin{equation}
\begin{aligned}
	\ln \Xi &= -128\pi^2 \int_{r_H +\epsilon}^{r_H +\epsilon +\delta} \frac{r^2}{f^3} \upd r \int_{0}^{\infty} \omega^2 \ln (1-e^{-\beta \hbar \omega}) \upd \omega \\
	&+1536\pi^2 \alpha \int_{r_H +\epsilon}^{r_H +\epsilon +\delta}  \frac{r^2}{f^5} \upd r \int_{0}^{\infty} \omega^4 \ln (1-e^{-\beta \hbar \omega}) \upd \omega.
\end{aligned}
\end{equation}
In the vicinity of the event horizon, expanding $f(r)$ with the Taylor series and taking the first approximation lead to
\begin{equation}
    f(r) =f(r_H) + f'(r_H)(r-r_H) + \cdots =f'(r_H)(r-r_H).
\end{equation}
Finally we obtained the concrete expression of the grand partition function
\begin{equation}
    \ln \Xi = \frac{32 \pi^5 V}{45 f'^3 \epsilon^3 \hbar^3 \beta^3} - \frac{1024\pi^{7} V \alpha}{105 f'^5 \epsilon^5 \hbar^5 \beta^5},
\end{equation}
where $V=4\pi r_H^2 \delta$. 
The internal energy $U$ and pressure $p$ of the Bose gas can be derived from the partition function as
\begin{equation}
\label{eq.U}
    U= - \frac{\partial \ln \Xi}{\partial \beta} = \frac{32 \pi^5 V k_B^4 }{15 f'^3 \epsilon^3 \hbar^3 }T^4 - \frac{1024\pi^{7} V k_B^6 }{21 f'^5 \epsilon^5 \hbar^5} \alpha T^6,
\end{equation}
\begin{equation}
    p =\frac{1}{\beta} \frac{\partial \ln \Xi}{\partial V} = \frac{32 \pi^5 k_B^4 }{45 f'^3 \epsilon^3 \hbar^3}T^4 - \frac{1024 \pi^{7} k_B^6 }{105 f'^5 \epsilon^5 \hbar^5}\alpha T^6.
\end{equation}
Comparing the two equations above, the relationship between $p$ and the internal energy density $u$ can be obtained as follows 
\begin{equation}
\label{eq.pu}
    p=\frac{1}{3}u +\frac{2048\pi^7 k_B^6}{315 f'^5 \epsilon^5 \hbar^5} \alpha T^6,
\end{equation}
where 
\begin{equation}
\label{eq.u}
    u=\frac{U}{V}= \frac{32 \pi^5 k_B^4 }{15 f'^3 \epsilon^3 \hbar^3}T^4 - \frac{1024\pi^{7} k_B^6 }{21 f'^5 \epsilon^5 \hbar^5}\alpha T^6.
\end{equation}
Also we can obtain the entropy of the Bose gas in the shell layer as
\begin{equation}
\label{eq.S}
    S= k_B(\ln \Xi +\beta U)= \frac{128 \pi^5 k_B^4 V}{45 f'^3 \epsilon^3 \hbar^3} T^3 - \frac{2048\pi^7 k_B^6 V}{35 f'^5 \epsilon^5 \hbar^5}\alpha T^5.
\end{equation}
Obviously, the second term of Eq.(\ref{eq.S}) is a correction, which is related to the GUP. Since $V=4\pi r_H^2 \delta=A \delta$, by taking a proper cutoff factor $\frac{\delta}{\epsilon^3}=\frac{45\hbar^3}{8\pi^2 k_B}$, the first term of Eq.(\ref{eq.S}) can be expressed as $A/4$, and the second term becomes $-\frac{9\alpha A}{28\hbar^2 \epsilon^2}$. In addition, if we let the corrected entropy in Eq.(\ref{eq.S}) equal $A/4$, then the cutoff factor $\epsilon$ will be a function of $\delta$, satisfying the equation $315\hbar^5 \epsilon^5-56\pi^2k_B \delta \epsilon^2+72\pi^2k_B \delta=0$.

\section{Discussions and conclusion}
\subsection{Effects of curved spacetime and GUP on the thermodynamics quantities}
As can be seen from the results above, we obtained two corrections. The first correction is the presentation of the gravitation effect in the coefficient terms. The curved spacetime causes the coefficients to change such that they all contain the zeroth component $f'$ of the metric. Secondly, the GUP is responsible for the emergence of the higher-order term. It follows that the influence of the two effects makes the thermodynamic quantities of Bose gas different from those in the flat spacetime. We already know that the Hawking temperature of a black hole is inversely proportional to the mass. Generally, the mass is so large that the temperature is very small. In contrast, $T^5$ and $T^6$ give a small amount of correction. However, as the mass decreases, the radiation temperature gets much higher and higher. In the final stage of radiation, $T^5$ and $T^6$ play a vital role. In Ref.\cite{He_2014}, He et al. showed that a black hole reaches the highest temperature when the energy flow density is zero, at which point the black hole will cease to radiate and leave a remnant. As for the maximum temperature, it is discussed in the following subsections.

\subsection{Equation of state}
In the classical case, taking the photon gas as an example, the relationship between the radiation pressure $p'$ and the internal energy density $u'$ is
\begin{equation}
    p'=\frac{1}{3}u'.
\end{equation}
But Eq.(\ref{eq.pu}) gives that the pressure $p$ has a correction for the $T^6$ term. It suggests that the equation of state for Bose gas differs from that of the flat spacetime under strong gravitation effects. As temperature rises, the $T^6$ becomes dominant. Thus it would give a large correction.

\subsection{The corrected Stefan-Boltzmann law}
The relationship between radiation flux energy and internal energy density is
\begin{equation}
    J_u=\frac{1}{4} u.
\end{equation}
Substituting Eq.(\ref{eq.u}) into the above equation, we have
\begin{equation}
\label{eq.Ju}
    J_u=\sigma \eta_1 T^4 -\sigma \eta_2 T^6,
\end{equation}
where
\begin{equation}
    \eta_1 =\frac{32\pi^3}{f'^3 \epsilon^3},
\end{equation}\begin{equation}
    \eta_2 =\frac{5120\pi^5 k_B^2 \alpha}{7f'^5 \epsilon^5 \hbar^2},
\end{equation}
$\sigma=\frac{\pi^2 k_B^4}{60\hbar^3}$ is the Boltzmann constant. Equation (\ref{eq.Ju}) is the corrected Stefan-Boltzmann law. When $J_u=0$, the black hole radiation ceases and we obtain the highest temperature as
\begin{equation}
    T_c=\sqrt{\frac{7}{160\alpha}} \frac{f' \epsilon \hbar}{\pi k_B}.
\end{equation}

\subsection{Physical quantities in extreme condition}
In the final stage of the Schwarzschild black hole radiation, the black hole mass $M$ is very small and the temperature $T$ is close to the maximum temperature. Then we have
\begin{equation}
    p_c=\frac{49\pi f' \epsilon \hbar}{90000\alpha^2},
\end{equation}
\begin{equation}
    J_u=0,
\end{equation}
\begin{equation}
    S_c=\frac{7}{2250}  \sqrt{\frac{7}{10}} (\frac{1}{\alpha})^{3/2} k_B \pi^2.
\end{equation}
From the above analysis, it is clear that when the maximum temperature is reached, neither the pressure $p_c$ nor entropy $S_c$ appears negative, however, the average energy flux of black hole is zero. When the maximum temperature is approached, we believe that the method fails, because the first-order approximation of Taylor expansion is taken in the process, but it may not converge at maximum temperature. Therefore the method is only applicable at lower temperatures.

\subsection{Conclusion}
Since the gravitation effect near the event horizon cannot be negligible, we first used the Plainlevé metric which behaves well at the horizon, and the relativistic Klein-Gordon equation to obtain the energy-momentum relationship (\ref{eq.EMrlt}) of Bose gas in curved spacetime. The relation (\ref{eq.EMrlt}) is related to the $g_{00}$ component of the Schwarzschild spacetime, that is to say, it is related to the curvature of spacetime. We then introduce the GUP, which also considers the gravitation effects, into the grand partition function of statistical theory. According to the results, the thermodynamic quantities of Bose gas all appear to have a higher-order correction term due to the GUP.

\acknowledgments
This research is supported by National Natural Science Foundation of China under Grant No.11873025.

\bibliographystyle{eplbib}
\bibliography{references}

\begin{thebibliography}{10}
\expandafter\ifx\csname url\endcsname\relax\def\url#1{\texttt{#1}}\fi

\bibitem{hawking1971gravitational}
\Name{Hawking S.~W.} \REVIEW{Phys. Rev. Lett.}{26}{1971}{1344}.

\bibitem{bekenstein2020black}
\Name{Bekenstein J.~D.} \REVIEW{Phys. Rev. D}{7}{1973}{2333}.

\bibitem{hawking1975particle}
\Name{Hawking S.~W.} \REVIEW{Commun. Math. Phys.}{43}{1975}{199}.

\bibitem{veneziano1986stringy}
\Name{Veneziano G.} \REVIEW{Europhys. Lett.}{2}{1986}{199}.

\bibitem{GROSS1988407}
\Name{Gross D.~J. \and Mende P.~F.} \REVIEW{Nucl. Phys. B}{303}{1988}{407}.

\bibitem{yoneya1989interpretation}
\Name{Yoneya T.} \REVIEW{Mod. Phys. Lett. A}{4}{1989}{1587}.

\bibitem{AMATI198941}
\Name{Amati D., Ciafaloni M. \and Veneziano G.} \REVIEW{Phys. Lett.
  B}{216}{1989}{41}.

\bibitem{KONISHI1990276}
\Name{Konishi K., Paffuti G. \and Provero P.} \REVIEW{Phys. Lett.
  B}{234}{1990}{276}.

\bibitem{MAGGIORE199365}
\Name{Maggiore M.} \REVIEW{Phys. Lett. B}{304}{1993}{65}.

\bibitem{kempf1995hilbert}
\Name{Kempf A., Mangano G. \and Mann R.~B.} \REVIEW{Phys. Rev.
  D}{52}{1995}{1108}.

\bibitem{SCARDIGLI199939}
\Name{Scardigli F.} \REVIEW{Phys. Lett. B}{452}{1999}{39}.

\bibitem{adler1999gravity}
\Name{Adler R.~J. \and Santiago D.~I.} \REVIEW{Mod. Phys. Lett.
  A}{14}{1999}{1371}.

\bibitem{ahluwalia2000wave}
\Name{Ahluwalia D.~V.} \REVIEW{Phys. Lett. A}{275}{2000}{31}.

\bibitem{adler2001generalized}
\Name{Adler R.~J., Chen P. \and Santiago D.~I.} \REVIEW{Gen. Rel.
  Grav.}{33}{2001}{2101}.

\bibitem{Cust_dio_2003}
\Name{Custdio P.~S. \and Horvath J.~E.} \REVIEW{Class. Quant.
  Grav.}{20}{2003}{L197}.

\bibitem{CHEN2005233}
\Name{Chen P.} \REVIEW{New Astron. Rev.}{49}{2005}{233}.

\bibitem{MAZIASHVILI2006232}
\Name{Maziashvili M.} \REVIEW{Phys. Lett. B}{635}{2006}{232}.

\bibitem{CHEN20151}
\Name{Chen P., Ong Y.~C. \and Yeom D.~H.} \REVIEW{Phys. Rept.}{603}{2015}{1}.

\bibitem{medved2004conceptual}
\Name{Medved A. \and Vagenas E.~C.} \REVIEW{Phys. Rev. D}{70}{2004}{124021}.

\bibitem{MYUNG2007393}
\Name{Myung Y.~S., Kim Y.~W. \and Park Y.~J.} \REVIEW{Phys. Lett.
  B}{645}{2007}{393}.

\bibitem{LI2007207}
\Name{Li X.} \REVIEW{Phys. Lett. B}{647}{2007}{207}.

\bibitem{kim2007entropy}
\Name{Kim W., Kim Y.~W. \and Park Y.~J.} \REVIEW{Phys. Rev.
  D}{75}{2007}{127501}.

\bibitem{nozari2007existence}
\Name{Nozari K. \and Sefidgar A.} \REVIEW{Gen. Rel. Grav.}{39}{2007}{501}.

\bibitem{Nouicer_2007}
\Name{Nouicer K.} \REVIEW{Class. Quant. Grav.}{24}{2007}{5917}.

\bibitem{Kim_2008}
\Name{Kim W. \and Oh J.~J.} \REVIEW{JHEP}{2008}{2008}{034}.

\bibitem{PARK2008698}
\Name{Park M.} \REVIEW{Phys. Lett. B}{659}{2008}{698}.

\bibitem{HAN2008121}
\Name{Han X., Li H. \and Ling Y.} \REVIEW{Phys. Lett. B}{666}{2008}{121}.

\bibitem{nozari2012natural}
\Name{Nozari K. \and Saghafi S.} \REVIEW{JHEP}{2012}{2012}{1}.

\bibitem{He_2014}
\Name{He T.~M., Yang J.~B. \and Zhang J.~Y.} \REVIEW{Chin. Phys.
  Lett.}{31}{2014}{080401}.

\bibitem{feng2016quantum}
\Name{Feng Z.~W., Li H.~L., Zu X.~T. \and Yang S.~Z.} \REVIEW{Eur. Phys. J.
  C}{76}{2016}{212}.

\bibitem{THOOFT1985727}
\Name{'t~Hooft G.} \REVIEW{Nucl. Phys. B}{256}{1985}{727}.

\bibitem{sannan1988heuristic}
\Name{Sannan S.} \REVIEW{General Relativity and Gravitation}{20}{1988}{239}.

\bibitem{painleve1921mecanique}
\Name{Painlev{\'e} P.} \REVIEW{C. R. Acad. Sci. (Paris)}{173}{1921}{677}.

\end{thebibliography}

\end{document}